\documentclass[%
 reprint,
 amsmath,amssymb,
 aps,
 superscriptaddress,
]{revtex4-1}

\usepackage{graphicx}%
\usepackage{dcolumn}%
\usepackage{bm}%
\usepackage{color}
\definecolor{myurlcolor}{rgb}{0,0,0.7}
\definecolor{myrefcolor}{rgb}{0.8,0,0}
\usepackage[unicode=true,pdfusetitle, bookmarks=false,bookmarksnumbered=false,
bookmarksopen=false, breaklinks=false,pdfborder={0 0 0},backref=false,
colorlinks=true, linkcolor=myrefcolor,citecolor=myurlcolor,urlcolor=myurlcolor]
{hyperref}

\usepackage[caption=false]{subfig}
\usepackage{bbm}

\newcommand{\ket}[1]{\mathinner{\lvert#1\rangle}}

\newcommand{\Ket}[1]{\left| #1 \right>}

\newcommand{\Ketbra}[2]{\left|#1\middle>\middle<#2\right|}

\definecolor{orange}{RGB}{255,165,0}
\definecolor{darkgreen}{RGB}{0, 170, 0}

\begin{document}

\preprint{APS/123-QED}

\title{Machine learning for long-distance quantum communication}

\author{Julius Walln\"ofer}
\affiliation{Department of Physics, Freie Universit\"at Berlin, Arnimallee 14, 14195 Berlin, Germany}
\affiliation{Institute for Theoretical Physics, University of Innsbruck, Technikerstra{\ss }e 21a, 6020 Innsbruck, Austria}%
 
\author{Alexey A. Melnikov}%
\affiliation{Institute for Theoretical Physics, University of Innsbruck, Technikerstra{\ss }e 21a, 6020 Innsbruck, Austria}%
\affiliation{Department of Physics, University of Basel, Klingelbergstrasse 82, 4056 Basel, Switzerland}
\affiliation{Valiev Institute of Physics and Technology, Russian Academy of Sciences, Nakhimovskii prospekt 36/1, 117218 Moscow, Russia}

\author{Wolfgang D\"ur}
\affiliation{Institute for Theoretical Physics, University of Innsbruck, Technikerstra{\ss }e 21a, 6020 Innsbruck, Austria}%

\author{Hans J. Briegel}
\affiliation{Institute for Theoretical Physics, University of Innsbruck, Technikerstra{\ss }e 21a, 6020 Innsbruck, Austria}%
\affiliation{Department of Philosophy, University of Konstanz, Fach 17, 78457 Konstanz, Germany}%

\date{\today}%

\begin{abstract}
Machine learning can help us in solving problems in the context big data analysis and classification, as well as in playing complex games such as Go. But can it also be used to find novel protocols and algorithms for applications such as large-scale quantum communication? Here we show that machine learning can be used to identify central quantum protocols, including teleportation, entanglement purification and the quantum repeater. These schemes are of importance in long-distance quantum communication, and their discovery has shaped the field of quantum information processing. However, the usefulness of learning agents goes beyond the mere re-production of known protocols; the same approach allows one to find improved solutions to long-distance communication problems, in particular when dealing with asymmetric situations where channel noise and segment distance are non-uniform. Our findings are based on the use of projective simulation, a model of a learning agent that combines reinforcement learning and decision making in a physically motivated framework. The learning agent is provided with a universal gate set, and the desired task is specified via a reward scheme. From a technical perspective, the learning agent has to deal with stochastic environments and reactions. We utilize an idea reminiscent of hierarchical skill acquisition, where solutions to sub-problems are learned and re-used in the overall scheme. This is of particular importance in the development of long-distance communication schemes, and opens the way for using machine learning in the design and implementation of quantum networks. 
\end{abstract}

\maketitle

\section{Introduction}\label{sec:intro}
Humans have invented technologies with transforming impact on society. One such example is the internet, which significantly influences our everyday life. The quantum internet~\cite{Kimble08, Wehner:qinternetreview} could become the next generation of such a world-spanning network, and promises applications that go beyond its classical counterpart. This includes e.g. distributed quantum computation, secure communication or distributed quantum sensing. Quantum technologies are now at the brink of being commercially used, and the quantum internet is conceived as one of the key applications in this context. Such quantum technologies are based on the invention of a number of central protocols and schemes, for instance quantum cryptography~\cite{ShorQKD,RennerPhd,ZhaoQKD,GottesmanQKD,LoQKD} and teleportation~\cite{teleportation}. Additional schemes that solve fundamental problems such as the accumulation of channel noise and decoherence have been discovered and have also shaped future research. This includes e.g. entanglement purification~\cite{bbpssw, dejmps, eppreview} and the quantum repeater~\cite{Br98} that allow for scalable long-distance quantum communication. These schemes are considered key results whose discovery represent breakthroughs in the field of quantum information processing. But to what extent are human minds required to find such schemes?

Here we show that many of these central quantum protocols can in fact be found using machine learning by phrasing the problem in a reinforcement learning (RL) framework~\cite{RusselNorvig2003,sutton1998reinforcement,wiering2012reinforcement}, the framework at the forefront of modern artificial intelligence~\cite{human2015mnih,mastering2016silver,silver2018general}. By using projective simulation (PS)~\cite{briegel2012projective}, a physically motivated framework for RL, we show that teleportation, entanglement swapping, and entanglement purification are found by a PS agent. We equip the agent with a universal gate set, and specify the desired task via a reward scheme. With certain specifications of the structure of the action and percept spaces, RL then leads to the re-discovery of the desired protocols. Based on these elementary schemes, we then show that such an artificial agent can also learn more complex tasks and discover long-distance communication protocols, the so-called quantum repeaters \cite{Br98}. The usage of elementary protocols learned previously is of central importance in this case. We also equip the agent with the possibility to call sub-agents, thereby allowing for a design of a hierarchical scheme~\cite{simon2016,simon2017skill} that offers the flexibility to deal with various environmental situations. The proper combination of optimized block actions discovered by the sub-agents is the central element at this learning stage, which allows the agent to find a scalable, efficient scheme for long-distance communication. We are aware that we make use of existing knowledge in the specific design of the challenges. Rediscovering existing protocols under such guidance is naturally very different from the original achievement (by humans) of conceiving of and proposing them in the first place, an essential part of which includes the identification of relevant concepts and resources. However, the agent does not only re-discover known protocols and schemes, but can go beyond known solutions. In particular, we find that in asymmetric situations, where channel noise and decoherence are non-uniform, the schemes found by the agent outperform human-designed schemes that are based on known solutions for symmetric cases.

From a technical perspective, the agent is situated in stochastic environments~\cite{RusselNorvig2003,sutton1998reinforcement,universal2013orseau}, as measurements with random outcomes are central elements of some of the schemes considered. This requires to learn proper reactions to all measurement outcomes, e.g., the required correction operations in a teleportation protocol depending on outcomes of (Bell) measurements. Additional elements are abort operations, as not all measurement outcomes lead to a situation where the resulting state can be further used. This happens for instance in entanglement purification, where the process needs to be restarted in some cases as the resulting state is no longer entangled. The overall scheme is thus probabilistic. These are new challenges that have not been treated in projective simulation before, but the PS agent can in fact deal with such challenges. Another interesting element is the usage of block actions that have been learned previously. This is a mechanism similar to hierarchical skill learning in robotics~\cite{simon2016,simon2017skill}, and to clip composition in PS~\cite{briegel2012projective,briegel2012creative,melnikov2018active}, where previously learned tasks are used to solve more complex challenges and problems. Here we use this concept for long-distance communication schemes. The initial situation is a quantum channel that has been subdivided by multiple repeater stations that share entangled pairs with their neighboring stations. Previously learned protocols, namely entanglement swapping and entanglement purification, are used as new primitives. Additionally, the agent is allowed to employ sub-agents that operate in the same way but deal with a problem at a smaller scale, i.e. they find optimized block actions for shorter distances that the main agent can employ at the larger scale.  This allows the agent to deal with big systems, and re-discover the quantum repeater with its favorable scaling. The ability to delegate is of special importance in asymmetric situations as such block actions need to be learned separately for different initial states of the environment -- in our case the fidelity of the elementary pairs might vary drastically either because they correspond to segments with different channel noise, or they are of different length. In this case, the agent outperforms human-designed protocols that are tailored to symmetric situations.

The paper is organized as follows. In Sec.~\ref{sec:RL} we provide background information on reinforcement learning and projective simulation, and discuss our approach on how to apply these techniques on problems in quantum communication. In Sec.~\ref{sec:elementary}, we show that the PS agent can find solutions to elementary quantum protocols, thereby re-discovering teleportation, entanglement swapping, entanglement purification and the elementary repeater cycle. In Sec.~\ref{sec:scaling} we present results for the scaling repeater in a symmetric and asymmetric setting, and summarize and conclude in Sec.~\ref{sec:outlook}.

\section{Projective Simulation for quantum communication tasks\label{sec:RL}}

\begin{figure*}[ht!]
	\centering
	\includegraphics[width=0.75\linewidth]{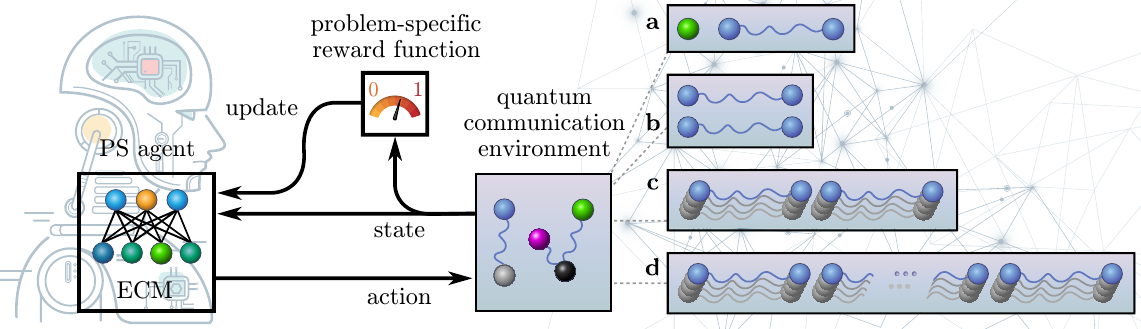}
	\caption{Illustration of the reinforcement learning agent interacting with the environment. The agent performs actions that change the state of the environment, while the environment communicates information about its state to the agent. The reward function is customized for each environment. The initial states for the different environments we consider here are illustrated: \textbf{a} Teleportation of an unknown state. \textbf{b} Entanglement purification applied recurrently. \textbf{c} Quantum repeater with entanglement purification and entanglement swapping. \textbf{d} Scaling quantum repeater concepts to distribute long-distance entanglement.
	}
	\label{fig:learning_scheme}
\end{figure*}

In this paper the process of designing quantum communication protocols is viewed as a reinforcement learning (RL) problem. RL, and more generally machine learning (ML), is becoming increasingly more useful in automation of problem-solving in quantum information science~\cite{dunjko2018machine,carleo2019machine,dunjko2020nonreview}. First, ML has been shown to be capable of designing new quantum experiments~\cite{melnikov2018active,driscoll2019quantum,krenn2020computer,melnikov2020setting} and new quantum algorithms~\cite{cincio2018learning,bang2014strategy}. Next, by bridging knowledge about quantum algorithms with actual near-term experimental capabilities, ML can be used to identify problems in which quantum advantage over a classical approach can be obtained~\cite{melnikov2019predicting,melnikov2020machinetransfer,moussa2020quantum}. Then, ML is used to realize these algorithms and protocols in quantum devices, by autonomously learning how to control~\cite{fosel2018reinforcement,xu2019generalizable,schafer2020differentiable}, error-correct~\cite{tiersch2015adaptive,nautrup2019optimizing,sweke2018reinforcement,agnes2019hamiltonian}, and measure quantum devices~\cite{liu2020repetitive}. Finally, given experimental data, ML can reconstruct quantum states of physical systems~\cite{shang2019reconstruction,carrasquilla2019reconstructing,torlai2019integrating}, learn a compact representation of these states and characterize them~\cite{carleo2017solving,gao2017efficient,canabarro2019machine}.

Here we propose learning quantum communication protocols by a trial and error process. This process is visualized in Fig.~\ref{fig:learning_scheme} as an interaction between an RL agent and its environment: by trial and error the agent is manipulating quantum states hence constructing communication protocols. At each interaction step the RL agent perceives the current state of the protocol (environment) and chooses one of the available operations (actions). This action modifies the previous version of the protocol and the interaction step ends. In addition to the state of the protocol the agent gets feedback at each interaction step. This feedback is specified by a reward function, which depends on the specific quantum communication task a)-d) in Fig.~\ref{fig:learning_scheme}. A reward is interpreted by the RL agent and its memory is updated.

The described RL approach is used for two reasons. First, there is a similarity between a target quantum communication protocol and a typical RL target. A target quantum communication protocol is a sequence of elementary operations leading to a desired quantum state, whereas a target of an RL agent is a sequence of actions that maximizes the achievable reward. In both cases the solution is therefore a sequence, which makes it natural to assign each elementary quantum operation a corresponding action, and to assign each desired state a reward. Second, the way the described targets are achieved is similar in RL and quantum communication protocols. In both cases an initial search (exploration) over a large number of operation (or action) sequences is needed. This search space can be viewed as a network, where states of a quantum communication environment are vertices, and basic quantum operations are edges. The structure of a complex network, formed in the described way, is similar to the one observed in quantum experiments~\cite{melnikov2018active}, which makes the search problem equivalent to navigation in mazes -- a reference problem in RL~\cite{sutton1998reinforcement,sutton1990integrated,mirowski2016learning,hierarchical2016maze}.

It should also be said, that the role of the RL agent goes beyond mere parameter estimation for the following reasons. First, using simple search methods (e.g, a brute-force or a guided search) would fail for the considered problem sizes: e.g. in the teleportation task discussed in section \ref{sec:teleportation}, the number of possible states of the communication environment is at least $7^{14} > 0.6\times 10^{12}$\footnote{In the teleportation task the shortest possible sequence of gates is equal to $14$, and at each step of this sequence there are at least $7$ possible gates that can be applied.}. 
Second, the RL agent learns in the space of its memory parameters, but it is not the case with optimization techniques (e.g, genetic algorithms, simulated annealing, or gradient descent algorithms) that would search directly in the parameter space of communication protocols. Optimizing directly in the space of protocols, which consist of both actions and stochastic environment responses, can only be efficient if the space is sufficiently small~\cite{sutton1998reinforcement}. Additional complication will be introduced by the fact that reward signals are often sparse in quantum communication tasks, hence the reward gradient is almost always zero giving optimization algorithms no direction for parameter change.
Third, using an optimization technique for constructing an optimal action sequence, ignoring stochastic environment responses, is usually not possible in quantum communication tasks. Because different responses are needed depending on measurement outcomes, there is no single action sequence that achieves an optimal protocol, i.e. there is no single point optimal point in the parameter space with which an optimization technique. Nevertheless, there is at least one point in the RL agent's memory parameter space that achieves an optimal protocol as the RL agent can choose an action depending on the current state of the environment rather than a whole action sequence.

As a learning agent that operates within the RL framework shown in Fig.~\ref{fig:learning_scheme} we use the PS agent~\cite{briegel2012projective,mautner2013projective}. PS is a physically-motivated approach to learning and decision making, which is based on deliberation in the episodic and compositional memory (ECM). The ECM is organized as an adjustable network of memory units, which provides flexibility in constructing different concepts in learning, e.g., meta-learning~\cite{makmal2016meta} and generalization~\cite{melnikov2017projective,ried2019minimal}. The deliberation within the ECM is based on a not computationally demanding random walk process, which in addition can be sped up via a quantum walk process~\cite{kempe2003quantum,venegas-andraca2012quantum}, leading to a quadratic speedup in deliberation time~\cite{paparo2014quantum,jerbi2019framework}, and makes the PS model conceptually attractive. Physical implementations of the basic PS agent and the quantum-enhanced PS agent were proposed by using photonic platforms~\cite{flamini2020photonic}, trapped ions~\cite{dunjko2015quantum}, and superconducting circuits~\cite{friis2015coherent}. The quantum-enhanced deliberation was recently implemented, as a proof-of-principle, in a small-scale quantum information processor based on trapped ions~\cite{sriarunothai2018speeding}.

\begin{figure*}[ht!]
	\centering
	\includegraphics[width=1\linewidth]{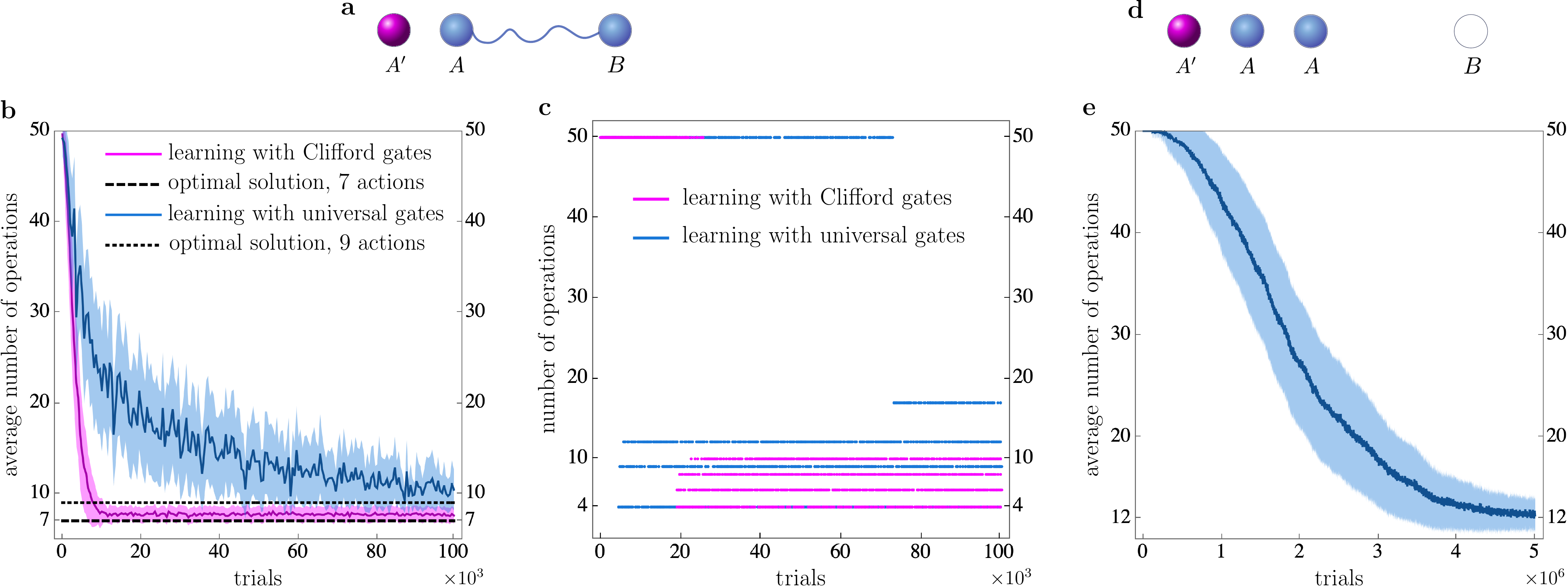}
	\caption{Reinforcement learning a teleportation protocol. \textbf{a} Initial setup: the agent is tasked to teleport the state of qubit $A^\prime$ to $B$ using a Bell pair shared between $A$ and $B$. \textbf{b} Learning curves of an ensemble of PS agents: average number of actions performed in order to teleport an unknown quantum state in the case of having Clifford gates (magenta) and universal gates (blue) as a set of available actions. \textbf{c} Two learning curves (magenta and blue) of two individual PS agents. Four solutions of different lengths are found by the agents. \textbf{d} Initial setup without pre-distributed entanglement. \textbf{e} Learning curve in the learning setting \textbf{d}: average number of actions performed in order to teleport an unknown quantum state. \textbf{b,e} The curves represent an average over $500$ agents. The shaded areas show mean squared deviation $\pm \sigma /3$. This deviation appears not only because of different individual histories of the agents, but also because of a difference in the individual solution lengths shown in \textbf{c}.}
	\label{fig:teleportation_results}
\end{figure*}

The use of PS in the design of quantum communication protocols has further advantages compared to other approaches, such as standard tabular RL models, or deep RL networks. First, the PS agent was shown to perform well on problems that, from an RL perspective, are conceptually similar to designing communication networks. In the problems that can be mapped to a navigation problem~\cite{melnikov2018benchmarking}, such as the design of quantum experiments~\cite{melnikov2018active} and the optimization of quantum error correction codes~\cite{nautrup2019optimizing}, PS outperformed methods that were practically used for those problems (and were not based on machine learning). In standard navigation problems, such as the grid world and the mountain car problem, the basic PS agent shows a performance qualitatively and quantitatively similar to standard tabular RL models of SARSA and Q-learning~\cite{melnikov2018benchmarking}. Second, as was shown in Ref.~\cite{melnikov2018benchmarking}, the computational effort is one to two orders of magnitude lower compared to tabular approaches. The reason for this is a low model complexity: in static task environments the basic PS agent has only one relevant model parameter. This makes it easy to set up the agent for a new complex environment, such as the quantum communication network, where model parameter optimization is costly because of the runtime of the simulations. Third, it has been shown that a variant of the PS agent converges to optimal behavior in a large class of Markov decision processes~\cite{clausen2019convergence}. Fourthly, by construction, the PS decision making can be explained by analyzing graph properties of its ECM. Because of this intrinsic interpretability of the PS model, we are able to properly analyze the outcomes of the learning process~\cite{melnikov2018active}.

Next, we show how the PS agent learns quantum communication protocols. The code of the PS agent used in this context is a derivative of a publicly available Python code~\cite{PScode}.

\section{Learning elementary protocols \label{sec:elementary}}

We let the agent interact with various environments where the initial states and goals correspond to well-known quantum information protocols. For each of the protocols we will first explain our formulation of the environment and the techniques we used. Then we discuss the solutions the agent finds before finally comparing them to the established protocols. A detailed description of the environments together with additional results can be found in the Appendix.

The learning process follows a similar structure for all of the environments as the agent interacts with the environment over multiple trials. One trial consists of multiple interactions between agent and environment. At the beginning of each trial the environment is initialized in the initial setup , so each individual trial starts again from a blank slate. The agent selects one of the available actions (which are specific to the environment), then the environment will provide information to the agent whether the goal has been reached (with a reward $R>0$) or not ($R=0$), together with the current percept. The agent then gets to choose the next action and this repeats until the trial ends either successfully, if the goal is reached, or unsuccessfully e.g. if a maximum number of actions is exceeded or if there are no more actions left. We call a successful sequence of actions the agent used in one trial a protocol.

\subsection{Quantum teleportation \label{sec:teleportation}}
The quantum teleportation protocol~\cite{teleportation} is one of the central protocols of quantum information. In the standard version, a maximally entangled state shared between two parties, $A$ and $B$, is used as a resource and serves to teleport the unknown quantum state of a third qubit, which is also held by party $A$, from $A$ to $B$. To achieve this, $A$ performs a Bell measurement, communicates the outcome to $B$ via classical communication, and then $B$ performs a correction operation (Pauli operation) depending on the measurement outcome. Notice that the same scheme serves for entanglement swapping when the qubit to be teleported is itself entangled with a fourth qubit held by another party.

\subsubsection{Basic protocol}

The agent is tasked to find a way to transmit quantum information without directly sending the quantum system to the recipient. As an additional resource a maximally entangled state shared between sender and recipient is available. The agent can apply operations from a (universal) gate set locally. This task challenges the agent without any prior knowledge to find the best (shortest) sequence of operations out of a large number of possible action sequences, which grows exponentially with a sequence length.

We describe the learning task as follows:
There are two qubits $A$ and $A^\prime$ at the sender's station and one qubit $B$ at the recipient's station. Initially, the qubits $A$ and $B$ are in a maximally entangled state $\Ket{\Phi^+} = \frac{1}{\sqrt{2}} \left( \Ket{00} + \Ket{11} \right)$ and $A^\prime$ is in an arbitrary input state $\ket{\Psi}$. The setup is depicted in Fig.~\ref{fig:teleportation_results}a. For this setup we consider two different sets of actions: The first is a Clifford gate set consisting of the Hadamard gate $H$ and the $P$-gate $P=\mathrm{diag}(1, i)$, as well as the controlled-NOT (CNOT) gate (which, as a multi-qubit operation, can only be applied on qubits at the same station, i.e. in this case only on $A$ and $A^\prime$). Furthermore, single-qubit measurements in the $Z$-basis are allowed with the agent obtaining a measurement outcome. The second set of actions replaces $P$ with $T=\mathrm{diag}(1, e^{i \pi /4})$, which results in a universal set of quantum gates. A detailed description of the actions and percepts can be found in Appendix \ref{sec:appendix_teleportation}.

The task is considered to be successfully solved if the qubit at $B$ is in state $\ket{\Psi}$. In order to ensure that this works for all possible input states, instead of using random input states, we make use of the Jamio{\l}kowski fidelity~\cite{jamiolkowski, jamfid} to evaluate if the protocol proposed by the agent is successful. This means we require that the overlap of the Choi-Jamio{\l}kowski state~\cite{jamiolkowski} $\Ket{\Phi^+}_{\widetilde{A}^\prime A^\prime}$ corresponding to the effective map generated by the suggested protocol with the Choi-Jamio{\l}kowski state corresponding to the optimal protocol is equal to $1$.

\begin{figure*}[ht!]
	\centering
	\includegraphics[width=0.9\linewidth]{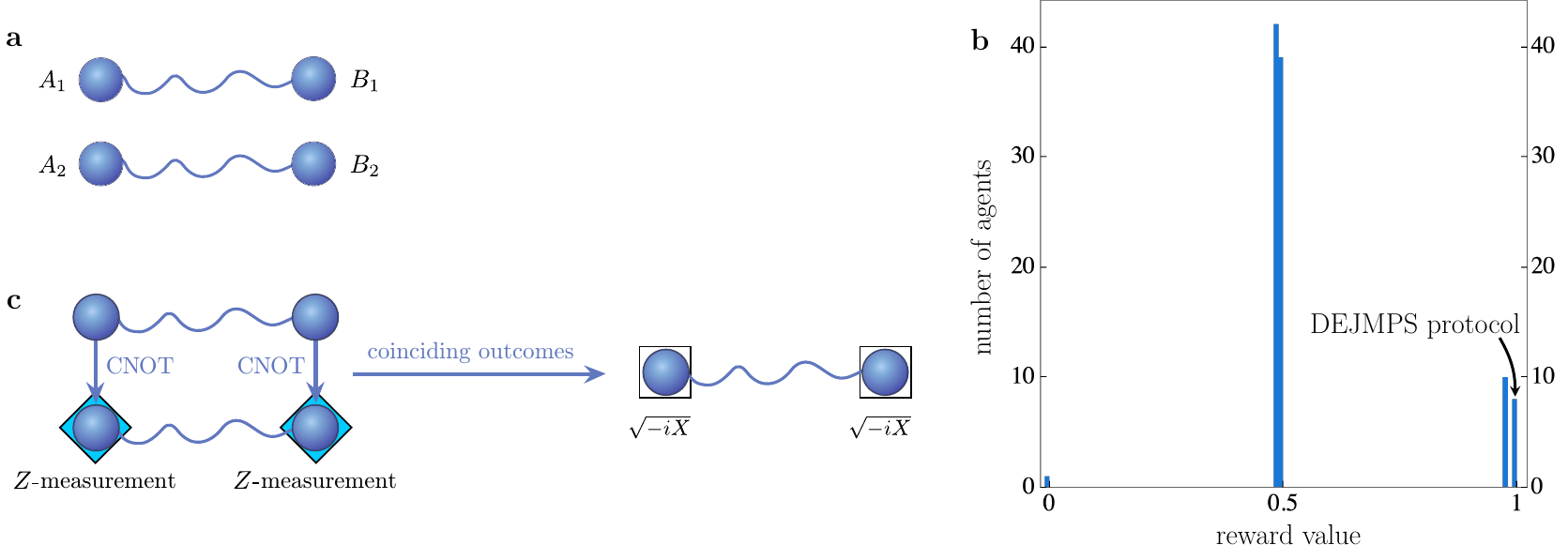}
	\caption{Reinforcement learning an entanglement purification protocol. \textbf{a} Initial setup of the quantum communication environment: two entangled noisy pairs shared between two stations $A$ and $B$. \textbf{b} Cumulative reward obtained by $100$ agents for their protocols found after $5 \times 10^5$ trials. \textbf{c} Illustration of the best protocol found by the agent: Apply bilateral $\mathrm{CNOT}$ operations and measure one of the pairs. If the measurement outcomes coincide, the protocol is considered successful and $\sqrt{-iX}$ is applied on both remaining qubits before the next entanglement purification step.
	}
	\label{fig:epp_results}
\end{figure*}

The learning curves, i.e the number of operations the agent applies to reach a solution at each trial, are shown as an average over $500$ agents in Fig.~\ref{fig:teleportation_results}b. Unsuccessful trials are recorded with $50$ operations as that is the maximum number of operations per trial to which the PS agent was limited. Observing the number of operations decrease below $50$ means that the PS agent finds a solution. The decline over time in the averaged learning curve stems not only from an increasing number of agents finding a solution but also from individual agents improving their solution based on their experience. We observe that the learning curve converges to some average number of operations in both cases, using a Clifford (magenta) and a universal (blue) gate set. However, the mean squared deviation does not go to zero. This can be explained by looking at the individual learning curves of two example agents in Fig.~\ref{fig:teleportation_results}c: the agent does not arrive at a single solution for this problem setup, but rather four different solutions. These solutions can be summarized as follows (up to different orders of commuting operations): 
\begin{itemize}
    \item Apply $\textrm{H}_{A^\prime} \textrm{CNOT}^{A^\prime \rightarrow A}$, where $\textrm{H}$ is the Hadamard gate and $\textrm{CNOT}$ is the controlled-NOT operation.
    \item Measure qubits $A$ and $A^\prime$ in the computational basis.
    \item Depending on the measurement outcomes, either apply $\mathbbm{1}$, $X$, $Y$ or $Z$ (decomposed to the elementary gates of the used gate set) on qubit $B$.
\end{itemize}
We see four different solutions in Fig.~\ref{fig:teleportation_results}c as four horizontal lines, which appear because of the probabilistic nature of the quantum communication environment. The agent learns different sequences of gates because different operations are needed, depending on measurement outcomes the agent has no control over. Four appropriate correction operations of different length (as seen in Fig.~\ref{fig:teleportation_results}c), which are needed in order for the agent to successfully transmit quantum information at each trial, complete the protocol. This protocol found by the agent is identical to the well-known quantum teleportation protocol~\cite{teleportation}.

Note that because we used the Jamio{\l}kowski fidelity to verify that the protocol implements the teleportation channel for all possible input states, it follows that the same protocol can be used for entanglement swapping if the input qubit at $A^\prime$ is part of an entangled state.

\subsubsection{Variants without pre-distributed entanglement}

The entangled state shared between two distant parties is the key resource that makes the quantum teleportation protocol possible. Naturally, one could ask whether the agent is still able to find a protocol if not provided with the initial entangled state. To this end we let the agent solve two variants of this task with the goal of transferring an input state $\Ket{\Psi}$ to the receiving station $B$ without sending it directly.
\paragraph{Variant 1:}
Initially there are two qubits $\Ket{0}_{A_1}\Ket{0}_{A_2}$ and the input qubit in state $\Ket{\Psi}$, all at the sender's station $A$. Note that in this variant there is no qubit at station $B$ initially. In addition to a universal gate set (multi-qubit operations can only be applied to qubits at the same station) the agent now has the additional capability to send a qubit to the recipient's station with the important restriction that the input qubit may not be sent.

For this case the agent quickly finds a solution, which is however different from the standard teleportation protocol and only uses one of the provided qubits $A_1$, $A_2$. The protocol is (up to order of commuting operators and permutations of qubits $A_1$ and $A_2$) given by: Apply $\textrm{H}_{A^\prime} \textrm{CNOT}^{A^\prime \rightarrow A_1}$, then send qubit $A_1$ to the receiving station. Now measure qubit $A^\prime$ in the computational basis. If the outcome is $-1$, apply the Pauli-$Z$ operator on qubit $A_1$. This protocol was called \textit{one-bit teleportation} in \cite{onebitTeleportation} and is conceptually and in terms of resources even simpler than the standard teleportation.

\paragraph{Variant 2:}
Now let us consider the same environment as in Variant 1 but with the additional restriction that no gates or measurements can be applied to the input qubit $A^\prime$ until one of the two other qubits has been sent to the recipient's station. In this case the agent indeed finds the standard quantum teleportation protocol, by first creating a Bell pair via $\textrm{CNOT}^{A_1 \rightarrow A_2} H_{A_1}$ and sending $A_2$ to the second station -- the rest is identical to the base case discussed before. In Fig.~\ref{fig:teleportation_results}e the learning curve averaged over 500 agents is shown. Obviously it takes significantly more trials for the agent to find solutions as the action sequence is longer. Nonetheless, this shows that the agent can find the quantum teleportation protocol without being given an entangled state as an initial resource.

\subsection{Entanglement purification \label{sec:epp}}

Noise and imperfections are a fundamental obstacle to distribute entanglement over long-distances, so a strategy to deal with these is needed. Entanglement purification is one approach that is integral to enabling long-distance quantum communication. It is a probabilistic protocol that generates out of two noisy copies of a (non-maximally) entangled state a single copy with increased fidelity. Iterative application of this scheme yields pairs with higher and higher fidelity, and eventually maximally entangled pairs are generated.

In particular here we investigate a situation that uses a larger amount of entanglement in the form of multiple noisy Bell pairs, each of which may have been affected by noise during the initial distribution, and try to obtain fewer, less noisy pairs from them. Again, the agent has to rely on using only local operations at the two different stations that are connected by the Bell pairs.

Specifically, we provide the agent with two noisy Bell pairs $\rho_{A_1B_1} \otimes \rho_{A_2B_2}$ as input, where $\rho$ is of the form of $\rho = F \Ketbra{\Phi^+}{\Phi^+} + \frac{1-F}{3} \left(\Ketbra{\Psi^+}{\Psi^+} + \Ketbra{\Phi^-}{\Phi^-} + \Ketbra{\Psi^-}{\Psi^-}\right)$. Here $\ket{\Phi^\pm}$ and $\ket{\Psi^\pm}$ denote the standard Bell basis and $F$ is the fidelity with respect to $\ket{\Phi^+}$. This starting situation is depicted in Fig.~\ref{fig:epp_results}a. The agent is tasked with finding a protocol that probabilistically outputs one copy with increased fidelity. However, it is desirable to obtain a protocol that does not only result in an increased fidelity when applied once, but consistently increases the fidelity when applied recurrently, i.e. on two pairs that have been obtained from the previous round of the protocol. In order to make such a recurrent application possible while dealing with probabilistic measurements, identifying the branches that should be reused is an integral part.

To this end, a different technique than before is employed. Rather than simply obtaining a random measurement outcome every time the agent picks a measurement action, instead the agent needs to provide potentially different actions for all possible outcomes. The actions taken on all the different branches of the protocol are then evaluated as a whole. This makes it possible to calculate the result of the recurrent application of that protocol separately for each trial. The agent is rewarded according to both the overall success probability of the protocol and the obtained increase in fidelity.

The agent is provided with a Clifford gate set and single-qubit measurements. Qubits labeled $A_i$ are held by one party and those labeled $B_i$ are held by another party. Multi-qubit operations can only be applied on qubits at the same station.
The output of each of the branches is enforced to be a state with one qubit on side $A$ and one on side $B$ along with a decision by the agent whether to consider that branch success or failure for the purpose of iterating the protocol. Since this naturally needs two single-qubit measurements, with two possible outcomes each, there are four branches that need to be considered.

\begin{figure}
	\centering
	\includegraphics[width=0.7\linewidth]{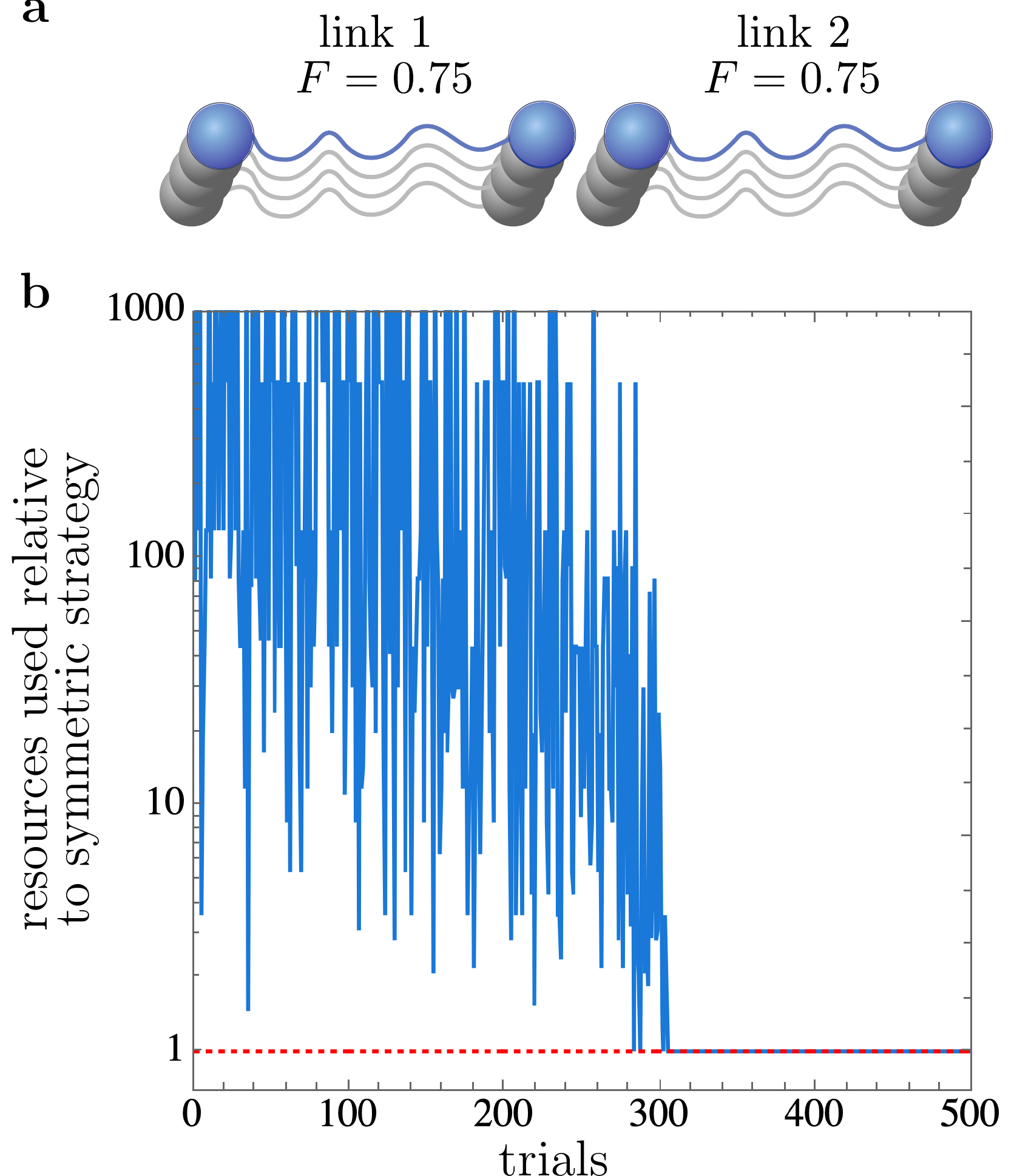}
	\caption{Reinforcement learning a quantum repeater protocol. \textbf{a}~Initial setup for the length 2 quantum repeater environment. The agent is provided with many copies of noisy Bell states with initial fidelities $F=0.75$, that can be purified on each link separately or connected via entanglement swapping at the middle station. \textbf{b}~Learning curve in terms of resources (used initial Bell pairs) for the best of 128 agents with gate reliability parameter $p=0.99$. The known repeater solution (red line) is reached.
	}
	\label{fig:repeater2_results}
\end{figure}

In Fig.~\ref{fig:epp_results}b we see reward values that $100$ agents obtained for the protocols applied to initial states with fidelities of $F=0.73$. The reward is normalized such that the entanglement purification protocol presented in Ref.~\cite{dejmps} would obtain a reward of $1.0$. All the successful protocols found start the same way (up to permutations of commuting operations): they apply $\mathrm{CNOT}^{A_1 \rightarrow A_2} \otimes \mathrm{CNOT}^{B_1 \rightarrow B_2}$ followed by measuring qubits $A_2$ and $B_2$ in the computational basis.
In some of the protocols two of the previously discussed four branches are marked as successful, while others only mark one particular combination of measurement outcomes. The latter therefore have a smaller probability of success, which is reflected in the reward. However, looking closely at the distribution in Fig.~\ref{fig:epp_results}b we can see that these cases correspond to two variants with slightly different rewards. Those variants differ in the operations that are applied on the output copies before the next purification step. The variant with slightly lower reward applies the Hadamard gate on both qubits: $H \otimes H$. The protocol that obtains the full reward of $1.0$ applies $\sqrt{-iX} \otimes \sqrt{-iX}$ and is depicted in Fig.~\ref{fig:epp_results}c. 
This protocol is equivalent to the well-known DEJMPS protocol~\cite{dejmps} for an even number of recurrence steps, but requires a shorter action sequence for the gate set provided to the agent. We discuss this solution in more detail, as well as an additional variant of the environment with automatic depolarization after each recurrence step, in Appendix~\ref{sec:appendix_epp}.

\begin{figure*}[ht!]
    \centering
    \includegraphics[width=1\linewidth]{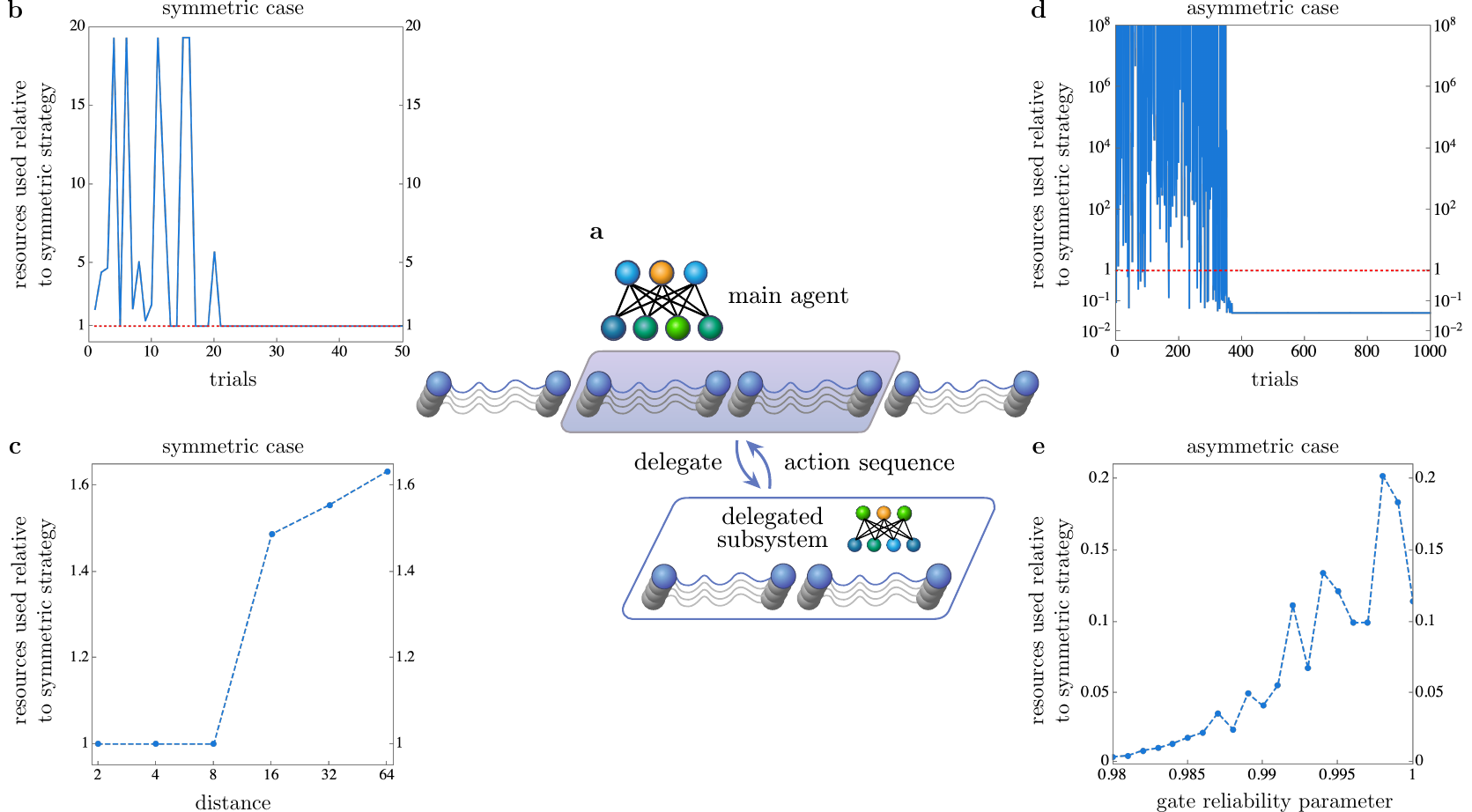}
    \caption{Reinforcement learning a scalable quantum repeater protocol. \textbf{a} Illustration of delegating a block action. The main agent that is tasked with finding a solution for the large scale problem (repeater length $4$ in this example) delegates the solution of a subsystem of length $2$ to another agent. That agent comes up with a solution for the smaller scale problem and that action sequence is then applied to the larger problem. This counts as one single action for the main agent. For settings even larger then this, the sub-agent itself can again delegate the solution of a sub-sub-system to yet another agent. \textbf{b-c} Scaling repeater with forced symmetric protocols with initial fidelities $F=0.75$. Gate reliability parameter $p=0.99$. The threshold fidelity for a successful solution is $0.9$. The red line corresponds to the solution with approximately $0.518\times 10^8$ resources used. \textbf{b} Best solution found by an agent with repeater length 8. \textbf{c} Relative resources used by the agent's solution compared to a symmetric strategy for different repeater lengths. \textbf{d-e} Scaling repeater with asymmetric initial fidelities $(0.8, 0.6, 0.8, 0.8, 0.7, 0.8, 0.8, 0.6)$. The threshold fidelity for a successful solution is $0.9$. \textbf{d} Best solution found by an agent with gate reliability $p=0.99$ outperforms a strategy that does not take the asymmetric nature of the initial state into account (red line). \textbf{e} Relative resources used by the agent's solution compared to a symmetric strategy for different reliability parameters\footnote{The jumps in the relative resources used are likely due to threshold effects or agents converging to a very short sequence of block actions that is not optimal.}. }
    \label{fig:scaling_repeater}
\end{figure*}

\subsection{Quantum repeater \label{sec:quantum_repeater}}

Entanglement purification alone certainly increases the distance over which one can distribute an entangled state of sufficiently high fidelity. However, the reachable distance is limited because at some point too much noise will accumulate such that the initial states will no longer have the minimal fidelity required for the entanglement purification protocol.  The insight at the heart of the quantum repeater protocol~\cite{Br98} is that one can split up the channels into smaller segments and use entanglement purification on short-distance pairs before performing entanglement swapping to create a long-distance pair. In the most extreme case with very noisy (but still purifiable) short-distance pairs the requirement of prior purification can easily be understood since entanglement swapping alone would produce a state that can no longer be purified, but this approach can also be beneficial when considering resource requirements at less extreme noise levels.

While the value of the repeater protocol lies in its scaling behavior, which becomes manifest when the number of links grows, for now the agent has to deal with only two channel segments that distribute noisy Bell pairs with a common station in the middle as depicted in Fig.~\ref{fig:repeater2_results}a. In this scenario the challenge for the agent is to use the protocols of the previous sections in order to distribute an entangled state over the whole distance. To this end the agent may use the previously discovered protocols for teleportation/entanglement swapping and entanglement purification as elementary actions, rather than individual gates.

The task is to find a protocol for distributing an entangled state between the two outer stations with a threshold fidelity of at least $0.9$, all the while using as few initial states as possible. The initial Bell pairs are considered to have initial fidelities of $F=0.75$. Furthermore, the CNOT gates used for entanglement purification are considered to be imperfect, which we model as local depolarizing noise with reliability parameter $p$ acting on the two qubits involved followed by the perfect CNOT operation \cite{eppreview}. The effective map $\mathcal{M}^{a \rightarrow b}_\mathrm{CNOT}$ is given by:

\begin{equation}
    \mathcal{M}_\mathrm{CNOT}^{a \rightarrow b}(p) \rho = \mathrm{CNOT}_{ab} \left( \mathcal{D}^{a}(p) \mathcal{D}^{b}(p) \rho \right) \mathrm{CNOT}_{ab}^\dagger
\end{equation}

where $\mathcal{D}^{i}(p)$ denotes the local depolarizing noise channel with reliability parameter $p$ acting on the $i$-th qubit:

\begin{equation}
    \mathcal{D}^{i}(p) \rho = p \rho + \frac{1-p}{4} \left( \rho + X^i \rho X^i + Y^i \rho Y^i + Z^i \rho Z^i \right)
    \label{eqn:ldn}
\end{equation}
with $X^i$, $Y^i$, $Z^i$ denoting the Pauli matrices acting on the $i$-th qubit.

While the point of such an approach only begins to show for much longer distances, which we take a look at in Sec. \ref{sec:scaling}, some key concepts can already be observed at small scales.

The agent naturally tends to find solutions that use a small number of actions in an environment that is similar to a navigation problem. However, this is not necessarily desirable here because the resources, i.e. the number of initial Bell pairs, is the figure of merit in this scenario rather than the number of actions. Therefore an appropriate reward function for this environment takes the used resources into account.

In Fig.~\ref{fig:repeater2_results}b the learning curve of the best of 128 agents in terms of resources used is depicted. Looking at the best solutions, the key insight is that it is beneficial to purify the short-distance pairs a few times before connecting them via entanglement swapping even though this way more actions need to be performed by the agent. This solution is in line with the idea of the established quantum repeater protocol~\cite{Br98}.

\begin{figure*}
    \centering
    \includegraphics[width=0.8\linewidth]{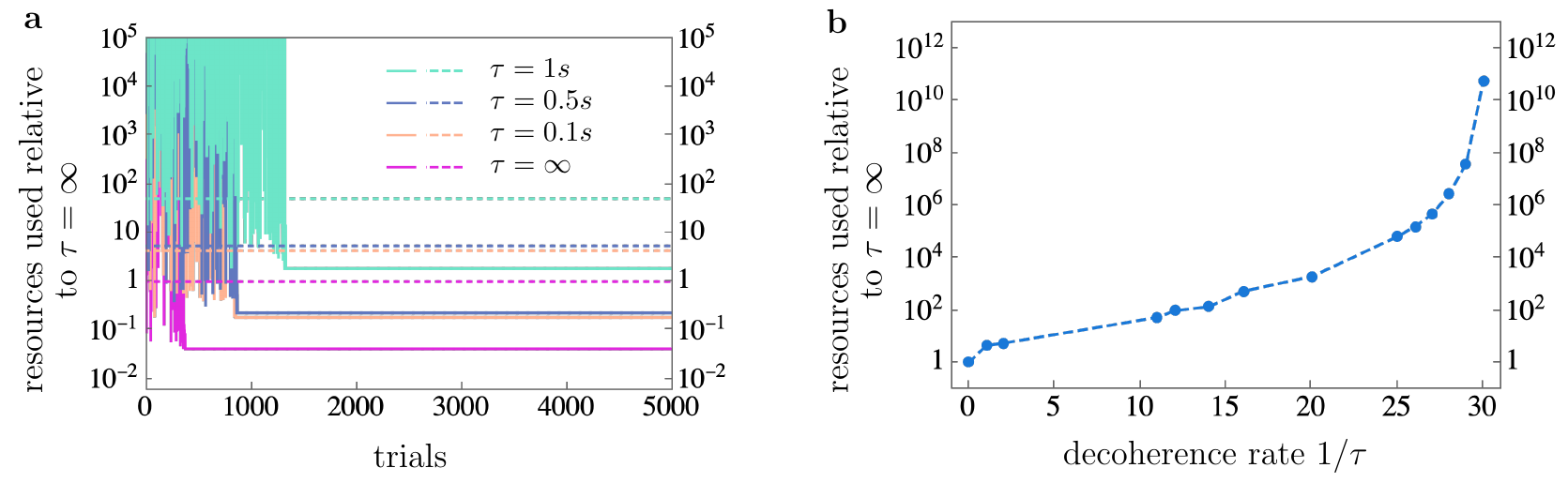}
    \caption{Reinforcement learning quantum repeater protocols with imperfect memories. We consider an asymmetric setup with repeater length $8$ and initial fidelities $(0.8, 0.6, 0.8, 0.8, 0.7, 0.8, 0.8, 0.6)$ arising from different distances between the repeater stations. Gate reliability parameter $p=0.99$. The threshold fidelity for a successful solution is $0.9$. \textbf{a} Learning curves (relative to the protocol designed for symmetric approaches for perfect memories) for different decoherence times $\tau$ of the quantum memories at repeater stations. The dashed lines are the resources used by symmetric strategies that do not account for the asymmetric situation.  \textbf{b} Required resources (number of initial pairs) of the protocols found by the agent for different memory times $\tau$ (relative to resources required for $\tau=\infty$).}
    \label{fig:fig6_memory_times}
\end{figure*}

\section{Scaling quantum repeater \label{sec:scaling}}

The point of the quantum repeater lies in its scaling behavior which only starts to show when considering longer distances than just two links. This means we have to consider starting situations of variable length as depicted in Fig.~\ref{fig:learning_scheme}d using the same error model as described in section \ref{sec:quantum_repeater}.
In order to distribute entanglement over varying distances, the agent needs to come up with a scalable scheme. However, both the action space and the length of action sequences required to find a solution would quickly grow unmanageable with increasing distances. Furthermore, an RL agent learns a solution for a particular situation and problem size rather than finding a universal concept that can be transferred to similar starting situations and larger scales.

To overcome these restrictions, we provide the agent with the ability to effectively \textit{outsource} finding solutions for distributing an entangled pair over a short distance and reuse them as elementary actions for the larger setting. This means that, as a single action, the agent can instruct multiple sub-agents to come up with a solution for a small distance and then pick the best action sequence among those solutions. This process is illustrated in Fig.~\ref{fig:scaling_repeater}a.

Again, the aim is to come up with a protocol that distributes an entangled pair over a long distance with sufficiently high fidelity, while using as few resources as possible.

\subsection{Symmetric protocols}
First, we take a look at a symmetric variant of this setup: The initial situation is symmetric and the agent is only allowed to do actions in a symmetric way. If it applies one step of an entanglement purification protocol on one of the initial pairs, all the other pairs need to be treated in the same way. Similarly, entanglement swapping is always performed at every second station that is still connected to other stations. In Fig.~\ref{fig:scaling_repeater}b-c the results for various lengths of Bell pairs with an initial fidelity of $F=0.75$ are shown. We compare the solutions that the agent found with a strategy that repeatedly purifies all pairs up to a chosen working fidelity followed by entanglement swapping (see Appendix~\ref{sec:wfstrategy}). For lengths greater than 8 repeater links, the agent still finds a solution with desirable scaling behavior solution while only using slightly more resources.

\subsection{Asymmetric setup}
The more interesting scenario is when the initial Bell pairs are subjected to different levels of noise, e.g. when the physical channels between stations are of different length or quality. In this scenario symmetric protocols are not optimal.

We consider the following scenario: 9 repeater stations connected via links that can distribute Bell pairs of different initial fidelities $(0.8, 0.6, 0.8, 0.8, 0.7, 0.8, 0.8, 0.6)$. In Fig.~\ref{fig:scaling_repeater}d the learning curve in terms of resources for the agent that can delegate work to sub-agents is shown. The gate reliability of the CNOT gates used in the entanglement purification protocol is $p=0.99$. The obtained solution is compared to the resources needed for a protocol that does not take into account the asymmetric nature of this situation and that is also used as an initial guess for the reward function (see Appendix \ref{sec:wfstrategy} for additional details of that approach). Clearly the solution found by the RL agent is preferable to the protocol tailored to symmetric situations. Fig.~\ref{fig:scaling_repeater}e shows how that advantage scales for different gate reliability parameters $p$.

\begin{table*}[htbp]
    \centering
    \begin{tabular}{c|ccc|ccc|c}
         station at & resources & \ \ \ \   & stations at & resources & \ \ \ \   & stations at & resources\\
         \hline
         4 & $9.60 \times 10^4$ & & 3, 6 & $1.06 \times 10^5$ & & 2, 4, 7 & $9.95 \times 10^4$ \\ 
         5 & $1.60 \times 10^5$ & & 3, 5 & $1.39 \times 10^5$ & & 3, 5, 7 & $1.07 \times 10^5$ \\
         3 & $1.60 \times 10^5$ & & 2, 5 & $1.59 \times 10^5$ & & 2, 4, 6 & $1.08 \times 10^5$ \\
         6 & $4.97 \times 10^5$ & & 3, 7 & $1.64 \times 10^5$ & & 1, 3, 6 & $1.10 \times 10^5$ \\
         2 & $5.02 \times 10^5$ & & 2, 6 & $1.64 \times 10^5$ & & 1, 3, 5 & $1.11 \times 10^5$ \\
         7 & $4.31 \times 10^6$ & & 4, 7 & $1.80 \times 10^5$ & & 2, 3, 5 & $1.11 \times 10^5$ \\
         1 & $5.90 \times 10^8$ & & 3, 4 & $1.90 \times 10^5$ & & 3, 4, 6 & $1.12 \times 10^5$ \\
    \end{tabular}
    \caption{Resource requirements for various choices to construct repeater stations in order to connect two distant parties that are $20 \text{\ km}$ apart. There are $7$ possible locations to place repeater stations that are asymmetrically spaced (see Appendix \ref{sec:appendix_repeater_positions}). Gate reliability parameter $p=0.99$, decoherence time for quantum memories $\tau=0.1 \text{\ s}$, attenuation length $L_\mathrm{att}=22 \text{\ km}$.}
    \label{tab:positions_table}
\end{table*}

\subsection{Imperfect memories \label{sec:imperfect_memories}}

One central parameter that influences the performance of the quantum repeater is the quality of quantum memories available at the repeater stations. It is necessary to store the qubits while the various measurement outcomes of the entanglement swapping and especially the entanglement purification protocols are communicated between the relevant stations.

To this end we revisit the previous asymmetric setup and assume that the initial fidelities now arise from different channel lengths between the repeater stations. We model the noisy channels as local depolarizing noise (see equation \eqref{eqn:ldn}) with length dependent error parameter $e^{-L/L_\mathrm{att}}$ with $L_\mathrm{att} = 22 \text{\ km}$ \footnote{This value is usually used for photon loss in a glass fibre, but here we use it for the decay in fidelity instead, which results in extremely noisy channels.}. Similarly, we model imperfect memories by local depolarizing noise with error parameter $e^{-t/\tau}$, where $t$ is the time for which a qubit is stored and $\tau$ is the decoherence time of the quantum memory.

The measurement outcomes of the entanglement purificaton protocol need to be communicated between the two stations performing the protocol. This information tells the stations whether the purification step was successful and is needed before the output pair can be used in any further operations. Therefore both qubits of the pair need to be stored for a time $t_\mathrm{epp}= l/c$, with the distance between repeater stations $l$ and the light speed in glass fibres $c = 2 \times 10^8 \mathrm{\ m/s}$. The measurement outcomes of the entanglement swapping also need to be communicated from the middle station performing the Bell measurement to the outer repeater stations, so that the correct by-product operator is known. In this case, however, it is not necessary to wait for this information to arrive before proceeding with the next operations. This is the case because all operations used in the protocols are Clifford operations thus any Pauli by-product operators can be taken into account retroactively. In our particular case, the information from entanglement swapping may only change the interpretation of the measurement outcomes obtained in the subsequent entanglement purification protocol and needs to be available at that time. However, since the information exchange for the entanglement purification protocol always happens over a longer distance than the associated entanglement swapping, this information will always be available when needed, so there is no need to account for additional time in memory for that.

Using the same approach as above, we let the agent learn protocols for different values of $\tau$ and compare them to the symmetric protocols. We use a gate reliability parameter $p=0.99$ and the task is considered complete if a pair with a fidelity $F > 0.9$ has been distributed. The learning curves as well as the advantage over the protocols optimized for symmetric approaches in Fig. \ref{fig:fig6_memory_times}a look qualitatively very similar to the result for perfect memories (i.e. $\tau = \infty$). In Fig. \ref{fig:fig6_memory_times}b the required initial pairs of the protocols found by the agent are shown for different memory times. The required resources increase sharply at a decoherence time of $\tau=1/30 \text{\ s}$ and the agent was unable to find a protocol for $\tau=1/31 \text{\ s}$ (which could either mean the threshold has been reached or the number of actions required for a successful protocol has grown so large that the agent could not find it in the a few thousand trials). It should be noted that this rather demanding memory requirement for this particular setup certainly arises from our very challenging starting situation (e.g. some very low starting fidelities of $0.6$).

\subsection{Choosing location of repeater stations \label{sec:finding_positions}}

As an alternative use case, the protocols found by the agent allow us to compare different setups. Let us consider the following situation: We want two distant parties to share entanglement via a quantum repeater by using only a small number of intermediate repeater stations. On the path between the terminal stations there are multiple possible locations where repeater stations could be built. Which combination of locations should be chosen? Furthermore, let us assume that the possible locations are unevenly spaced so that simply picking a symmetric setup is impossible.

We demonstrate this concept with the following example setup using the same error model as in Section \ref{sec:imperfect_memories} for both the length dependent channel noise and imperfect memories ($\tau=0.1 \text{\ s}$). We consider possible locations (numbered 1 to 7) for stations between the two end points that are located at positions that correspond to the asymmetric setup from the previous subsection but scaled down to a total distance of $L_\mathrm{tot} = 20 \text{\ km}$. The positions of all locations are listed in the appendix.
In Table \ref{tab:positions_table} we show the best combinations of locations for placing either 1, 2 or 3 repeater stations and the amount of resources the agent's protocol would take for these choices.

Naturally, this analysis could be repeated for different initial setups and error models of interest. This particular example can be understood as a proof-of-principle that using the agent in this way can be useful as well.

\section{Summary and Outlook \label{sec:outlook}}

We have demonstrated that reinforcement learning can serve as a highly versatile and useful tool in the context of quantum communication. When provided with a sufficiently structured task environment including an appropriately chosen reward function, the learning agent will retrieve (effectively re-discover) basic quantum communication protocols like teleportation, entanglement purification, and the quantum repeater. We have developed methods to state challenges that occur in quantum communication as RL problems in a way that offers very general tools to the agent while ensuring that relevant figures of merit are optimized.

We have shown that stating the considered challenges as an RL problem is beneficial and offers advantages over using optimization techniques as discussed in section~\ref{sec:RL}.

Regarding the question to what extent programs can help us in finding genuinely new schemes for quantum communication, it has to be emphasized that a significant part of the work consists in asking the right questions and identifying the relevant resources, both of which are central to the formulation of the task environment and are provided by researchers. However, it should also be noted that not every aspect of designing the environment is necessarily a crucial addition and many details of the implementation are simply an acknowledgment of practical limitations like computational runtimes. When provided with a properly formulated task, a learning agent can play a helpful, assisting role in exploring the possibilities.

In fact, we used the PS agent in this way to demonstrate that the application of machine learning techniques to quantum communication is not limited to rediscovering existing protocols. The PS agent finds adapted and optimized solutions in situations that lack certain symmetries assumed by the basic protocols, such as the qualities of physical channels connecting different stations.
We extended the PS model to include the concept of delegating parts of the solution to other agents, which allows the agent to effectively deal with problems of larger size. Using this new capability for long-distance quantum repeaters with asymmetrically distributed channel noise the agent comes up with novel and practically relevant solutions.

We are confident that the presented approach can be extended to more complex scenarios. We believe that reinforcement learning can become a practical tool to apply to quantum communication problems that do not have a rich spectrum of existing protocols such as designing quantum networks, especially if the underlying network structure is irregular. Alternatively, machine learning could be used to investigate other architectures for quantum communication such as constructing cluster states for all-photonic quantum repeaters \cite{Azuma2015}.

\section*{Acknowledgments}
J.W. and W.D. were supported by the Austrian Science Fund (FWF) through Grants No. P28000-N27 and P30937-N27. 
J.W. acknowledges funding by Q.Link.X from the BMBF in Germany.
A.A.M. acknowledges funding by the Swiss National Science Foundation (SNSF), through the Grant PP00P2-179109 and by the Army Research Laboratory Center for Distributed Quantum Information via the project SciNet.
H.J.B. was supported by the FWF through the SFB BeyondC F7102 and by the Ministerium f\"{u}r Wissenschaft, Forschung, und Kunst Baden-W\"{u}rttemberg (AZ: 33-7533.-30-10/41/1). 

\bibliographystyle{ieeetr}
\bibliography{psqnet.bib}

\clearpage
\appendix

\section{Introduction to projective simulation} 

\begin{figure}
    \centering
    \includegraphics[width=1\linewidth]{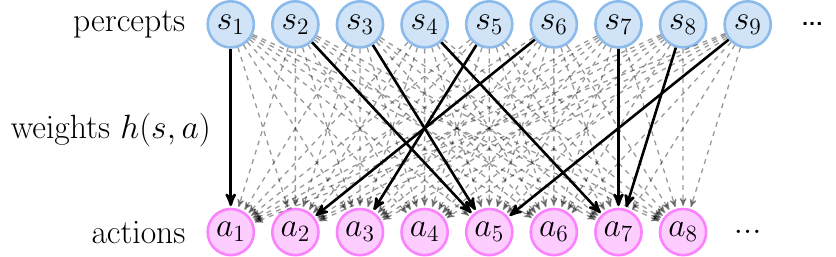}
    \caption{The PS network in its basic form, which corresponds to a weighed directed bipartite graph. Clips corresponding to $8$ states of environment and $7$ possible actions are shown. They are connected by directed edges. Thick edges correspond to high probabilities of choosing action $a_k$ in the state $s_m$.}
    \label{fig:PS_network}
\end{figure}

This work is based on using reinforcement learning (RL) for quantum communication. As an RL model we use the model of projective simulation (PS), which was first introduced in Ref.~\cite{briegel2012projective}. In the main text we provide background information on RL and PS, and explain our motivation of using PS. In this section we give an introduction to the working principles of the PS agent. The decision-making of a PS agent is realized in its episodic and compositional memory, which is a network on memory units, or clips. Each clip encodes a percept, an action, or a sequence thereof. There are different mechanisms to connect clips in the PS network, which can be found e.g. in Refs.~\cite{melnikov2018benchmarking,melnikov2017projective}. In this work we use a two-layered PS network construction, similar to the one used for designing quantum experiments~\cite{melnikov2018active}. The two-layered network is schematically shown in Fig.~\ref{fig:PS_network}.

The first layer of clips corresponds to percepts $s$, which are states of the (quantum communication) environment. The layer of percepts is connected to the layer of actions $a$ by edges with the weights $h(s,a)$. These weights determine the probabilities of choosing the corresponding action: the agent perceives a state $s_m$ and performs action $a_k$ with probability
\begin{equation}
    p_{mk} = \frac{\mathrm{e}^{h_{mk}}}{\sum_l \mathrm{e}^{h_{ml}}}.
\end{equation}
The decision-making process within the PS model, using the two-layered network, is hence a $1$-step random walk process. One trial consisting of $n$ agent-environment interaction steps is an $n$-step random walk process defined by the weight matrix $h_{mk}$. The weights of the network are updated at the end of each agent-environment interaction step $i$ according to the learning rule
\begin{equation}
    h^{(i+1)}_{mk} = h^{(i)}_{mk} - \gamma\left(h^{(i)}_{mk}-1\right) + g^{(i+1)}_{mk}r^{(i)},
\end{equation}
for all $m$ and $k$. $r^{(i)}$ is the reward, and $g^{(i+1)}$ is a coefficient that distributes this reward proportional to how much a given edge $(m,k)$ contributed to the rewarded sequence of actions. To be more specific, $g^{(i+1)}$ is set to $1$ once the edge is used in a random walk process, and goes back to it's initial value of $0$ with the rate $\eta$ afterwards:
\begin{equation}
g^{(i+1)}_{mk} = \begin{cases} 1, \mbox{ if } (m, k)\mbox{ was traversed} & \\
\left(1-\eta\right)g^{(i)}_{mk}, \mbox{ otherwise}. &
\end{cases}
\label{eq:glowRule}
\end{equation}
The time-independent parameter $\eta$ is set to a value from the interval $[0,1]$. The second meta-parameter $0\leq \gamma \leq 1$ of the PS agent is a damping parameter that helps the agent to forget, which is beneficial in cases where the environment changes.

\section{Quantum teleportation \label{sec:appendix_teleportation}}

\subsection{Environment description}
Fig.~\ref{fig:teleportation_results}a depicts the setup. Qubit $A^\prime$ is initialized as part of an entangled state $\ket{\Phi^+}_{\widetilde{A}^\prime A^\prime}$ in order to facilitate measuring the Jamio{\l}kowski fidelity later on. Qubits $A$ and $B$ are in a $\ket{\Phi^+}_{AB}$ state.

 \textbf{Goal:} The state of $A^\prime$ should be teleported to $B$. We measure this by calculating the Jamio{\l}kowski fidelity of the effective channel applied by the action sequences. That means we calculate the overlap of $\Ket{\Phi^+}_{\widetilde{A}^\prime B}$ with the reduced state $\rho_{\widetilde{A}^\prime B}$ to determine whether the goal has been reached.
 
 \textbf{Actions:} The following Actions are allowed:
    \begin{itemize}
     \item Depending on the specification of the task either $P = \begin{pmatrix} 1 & 0 \\
                            0 & i
                \end{pmatrix}$ for a Clifford gate set or 
                $T = \begin{pmatrix} 1 & 0 \\
                            0 & e^{i \pi / 4}
                \end{pmatrix}$ for a universal gate set, on each qubit. (3 actions)
     \item The Hadamard gate $H = \frac{1}{\sqrt{2}} \begin{pmatrix} 1 & 1 \\
                                                1 & -1
                                    \end{pmatrix}$ on each qubit (3 actions)
                                    
     \item The CNOT-gate $\mathrm{CNOT}^{A^\prime \rightarrow A}$  on the two qubits at location A. (1 action)
     \item Z-measurement on each qubit. (3 actions)
    \end{itemize}
    In total: \textbf{10 actions}. Note that the Clifford group is generated by $P$, $H$ and $CNOT$ \cite{CliffordGenerators} and replacing $P$ with $T$ makes it a universal gate set \cite{universalgateset}.
    The measurements are modeled as destructive measurements, which means that operations acting on that qubit are no longer available, thereby reducing the number of actions that the agent can choose.
    
    \textbf{Percepts:} The agent only uses the previous actions of the current trial as a percept. Z-measurements with different outcomes will produce different percepts.
    
    \textbf{Reward:} If Goal is reached, $R=1$ and the trial ends. Else $R=0$.

\subsection{Discussion}
One central complication in this scenario is that the entanglement is a limited resource. If the entanglement is destroyed without accomplishing anything (e.g., measuring qubit $A$ as the first action), then the goal can no longer be reached no matter what the agent tries afterwards. This is a feature that distinguishes this setup and other quantum environments with irreversible operations from a simple navigation problem in a maze. Instead this is more akin to a navigational problem, where there are numerous cliffs that the agent can fall down, but never get back up again, which means that the agent could be permanently separated from the goal.

An alternative formulation that would make the goal reachable at each trial even if a wrong irreversible action was taken would be to provide the agent with a reset action that resets the environment to the initial state. A different class of percepts would need to be used in this case.

With prior knowledge of the quantum teleportation protocol it is easy to understand why this problem structure favors an RL approach. The shortest solution for one particular combination of measurement outcomes takes only four actions and these actions are to be performed regardless of which correction operation needs to be applied. That means once this simple solution is found it significantly reduces the complexity of finding the other solutions, as now only the correction operations need to be found.

Compare this to searching for this solution by brute forcing action sequences. For the universal gate set we know that the most complicated of the four solutions takes at least $14$ actions. Ignoring the measurement actions for now as they anyway reduce the number of available actions, there are $7^{14}$ possible action sequences. So, we would have to try \textit{at least} $7^{14} > 6.7 \times 10^{11}$ sequences, which is vastly more than the few hundred thousands of trials needed by the agent.

\subsection{Environment variants without pre-distributed entanglement}
\paragraph{Variant 1:}
Fig.~\ref{fig:teleportation_results}d the initial setup is shown. The qubits $A_1$ and $A_2$ are both intialized in the state $\Ket{0}$. As before the input qubit $A^\prime$ is initialized as part of an entangled state $\ket{\Phi^+}_{\widetilde{A}^\prime A^\prime}$ in order to later obtain Jamio{\l}kowski fidelity.

\textbf{Goal:} A qubit at location $B$ is in the initial state of qubit $A^\prime$. As before the Jamio{\l}kowski fidelity is used to determine whether this goal has been reached.

\textbf{Actions:} The following actions are available:
\begin{itemize}
    \item The $T$-gate $T = \begin{pmatrix} 1 & 0 \\
                            0 & e^{i \pi / 4}
                \end{pmatrix}$ on every qubit (3 actions)
    \item The Hadamard gate $H = \frac{1}{\sqrt{2}} \begin{pmatrix} 1 & 1 \\
                                                1 & -1
                                    \end{pmatrix}$ on each qubit (3 actions)
    \item CNOT gate on each pair of qubits as long as they are at the same location (initially $\mathrm{CNOT}^{A^\prime \rightarrow A_1}$, $\mathrm{CNOT}^{A^\prime \rightarrow A_2}$ and $\mathrm{CNOT}^{A_1 \rightarrow A_2}$).
    \item Z-measurement on each qubit (3 actions).
    \item Send a qubit $A_1$ or $A_2$ to location $B$. (2 actions)
\end{itemize}
Initially available: \textbf{14 actions.} Measuring a qubit will remove all actions that involve that qubit from the pool of available actions. Sending a qubit to location $B$ will remove the used action itself (as that qubit is now at location $B$) and will also trigger a check which CNOT actions are now possible.

\textbf{Percepts:} The agent only uses the information which of the previous actions of the current trial wer taken as a percept. Z-measurements with different outcomes will produce different percepts.
    
\textbf{Reward:} If Goal is reached, $R=1$ and the trial ends. Else $R=0$.

\paragraph{Variant 2:}
As above, but initially no action involving the input qubit $A^\prime$ is available, i.e. no single qubit gates, no measurement nor CNOT gate. After one of the sending actions is used, these actions on qubit $A^\prime$ becom available. Furthermore, only one qubit may be sent in total, i.e. after sending a qubit neither of the two sending actions may be chosen.

\paragraph{Comment on the variant environments:}
Variant 1 serves as a good example why specifying the task correctly is such an important part for RL problems. As discussed in the main text, the agent immediately spotted the loophole and found a protocol that uses fewer actions than the standard teleportation protocol. Another successful way to circumvent the restriction of not sending $A^\prime$ directly is to construct a SWAP operation from the gate set. Then it is possible to simply swap the input state with one of the state of one of the qubits that can be sent. However, in the given action set this solution consists of a longer sequence of actions is therefore deemed more expensive than the one the agent found.

\section{Entanglement purification \label{sec:appendix_epp}}

\subsection{Environment description}
Fig.~\ref{fig:epp_results}a shows the initial setup with $A$ and $B$ sharing two Bell pairs with initial fidelity $F=0.73$.

\textbf{Goal:} Find an action sequence that results in a protocol that improves the fidelity of the Bell pair, when applied recurrently. This means that two copies of the resulting two-qubit state after one successful application of the protocol are taken and the protocol is using them as input states. 

\textbf{Actions:} The following Actions are available:
    \begin{itemize}
     \item $P_x = HPH$ on each qubit (4 actions)
     \item $H$, the Hadamard gate on each qubit (4 actions)
                                    
     \item The CNOT-gates $\mathrm{CNOT}^{A_1 \rightarrow A_2}$ and $\mathrm{CNOT}^{B_1 \rightarrow B_2}$  on qubits at the same location. (2 action)
     \item Z-measurements on each qubit (4 actions)
     \item Accept/Reject (2 actions)
    \end{itemize}
In total: \textbf{16 actions}. Note that these gates generate the Clifford group. (We tried different variants of generating the gate set as the choice of basis is not fundamental, the one with $P_x$ gave the best results.) The measurements are modeled as destructive measurements, which means that operations acting on that qubit are no longer available, thereby reducing the number of actions that the agent can choose. In order to reduce the huge action space further, the requirement that the final state of one sequence of gates needs to be a two-qubit state shared between $A$ and $B$ is enforced by removing actions that would destructively measure all qubits on one side. The accept and reject actions are essential because they allow identifying successful branches.

\textbf{Percepts:} The agent only uses the previous actions of the current trial as a percept. Z-measurements with different outcomes will produce different percepts.

\textbf{Reward:} The protocol suggested by the agent is performed recurrently for ten times. This is done to ensure that the solution found is a viable protocol for recurrent application because it is possible that a single step of the protocol might increase the fidelity but further applications of the protocol could undo that improvement. The reward function is given by $R=\mathrm{max}\left(0, \mathrm{const} \times \sqrt[10]{\prod_{i=1}^{10} p_i} \Delta F\right)$ where $p_i$ is the success probability (i.e. the combined probability of the accepted branches) of the $i$-th step, $\Delta F$ is the increase in fidelity after ten steps and the constant is chosen such that the known protocols \cite{bbpssw} or \cite{dejmps} would receive a reward of $1$. 

\textit{Problem-specific techniques:} To evaluate the performance of an entanglement purification protocol that is applied in a recurrent fashion it is necessary to know which actions are performed and especially whether the protocol should be considered successful for all possible measurement outcomes. Therefore, it is not sufficient to use the same approach as for the teleportation challenge and simply consider one particular measurement outcome for each trial. Instead, the agent is required to choose actions for all possible measurement outcomes every time it chooses a measurement action. This means we keep track of multiple separate branches (and the associated probabilities) with different states of the environment. The average density matrix of the branches that the agent decides to keep is the state that is used for the next purification step. We choose to do it this way because it allows us to obtain a complete protocol that can be evaluated at each trial and the agent is rewarded according to the performance of the whole protocol.

\subsection{Discussion}

As discussed in the main text, the agent found an entanglement purification protocol that is equivalent to the DEJMPS protocol \cite{dejmps} for an even number of purification steps.

Let us briefly recap how the DEJMPS protocol works:
Initially we have two copies of a state $\rho$ that is diagonal in the Bell basis and can be written with coefficients $\lambda_{ij}$:
\begin{equation}
    \begin{aligned}
        \rho = &\lambda_{00} \Ketbra{\Phi^+}{\Phi^+} + \lambda_{10} \Ketbra{\Phi^-}{\Phi^-} + \\
         &\lambda_{01} \Ketbra{\Psi^+}{\Psi^+} + \lambda_{11} \Ketbra{\Psi^-}{\Psi^-}
    \end{aligned}
\end{equation}

The effect of the multilateral CNOT operation $\mathrm{CNOT}^{A_1 \rightarrow A_2} \otimes \mathrm{CNOT}^{B_1 \rightarrow B_2}$ followed by measurements in the computational basis on $A_2$ and $B_2$ and post-selected for coinciding measurement results is:
\begin{equation}
    \begin{aligned}
        \widetilde{\lambda}_{00} &= \frac{\lambda^2_{00} + \lambda^2_{10}}{N} &
        \widetilde{\lambda}_{10} &= \frac{2 \lambda_{00} \lambda_{10}}{N} \\
        \widetilde{\lambda}_{01} &= \frac{\lambda^2_{01} + \lambda^2_{11}}{N} &
        \widetilde{\lambda}_{11} &= \frac{2 \lambda_{01} \lambda_{11}}{N} 
        \label{eqn:mcnot_only_map}
    \end{aligned}
\end{equation}
where $\widetilde{\lambda}_{ij}$ denote the new coefficient after the procedure and $N = \left(\lambda_{00} + \lambda_{10} \right)^2 + \left(\lambda_{01} + \lambda_{11} \right)^2$ is a normalization constant and also the probability of success. Without any additional intervention applying this map recurrently, not only the desired coefficient $\lambda_{00}$ (the fidelity) will be amplified, but both $\lambda_{00}$ and $\lambda_{10}$.

To avoid this and only amplify the fidelity with respect to $\ket{\Phi^+}$, the DEJMPS protocol calls for the application of $\sqrt{-iX} \otimes \sqrt{iX}$ on both copies of $\rho$ before applying the multilateral CNOTs and performing the measurements. The effect of this operation is to exchange the two coefficients $\lambda_{10}$ and $\lambda_{11}$ thus preventing the unwanted amplification of $\lambda_{10}$. So the effective map at each entanglement purification step is the following:
\begin{equation}
    \begin{aligned}
        \widetilde{\lambda}_{00} &= \frac{\lambda^2_{00} + \lambda^2_{11}}{N} &
        \widetilde{\lambda}_{10} &= \frac{2 \lambda_{00} \lambda_{11}}{N} \\
        \widetilde{\lambda}_{01} &= \frac{\lambda^2_{01} + \lambda^2_{10}}{N} &
        \widetilde{\lambda}_{11} &= \frac{2 \lambda_{01} \lambda_{10}}{N} 
        \label{eqn:DEJMPS_map}
    \end{aligned}
\end{equation}
with $N = \left(\lambda_{00} + \lambda_{11} \right)^2 + \left(\lambda_{01} + \lambda_{10} \right)^2$.

In contrast, the solution found by the agent calls for $\sqrt{-iX} \otimes \sqrt{-iX}$ to be applied, which exchanges two different coefficients $\lambda_{00}$ and $\lambda_{01}$ instead for an effective map:
\begin{equation}
    \begin{aligned}
        \widetilde{\lambda}_{00} &= \frac{\lambda^2_{01} + \lambda^2_{10}}{N} &
        \widetilde{\lambda}_{10} &= \frac{2 \lambda_{01} \lambda_{10}}{N} \\
        \widetilde{\lambda}_{01} &= \frac{\lambda^2_{00} + \lambda^2_{11}}{N} &
        \widetilde{\lambda}_{11} &= \frac{2 \lambda_{00} \lambda_{11}}{N} 
        \label{eqn:agent_epp_map}
    \end{aligned}
\end{equation}
and $N = \left(\lambda_{01} + \lambda_{10} \right)^2 + \left(\lambda_{00} + \lambda_{11} \right)^2$. Note that the maps \eqref{eqn:DEJMPS_map} and \eqref{eqn:agent_epp_map} are identical except that roles of $\widetilde{\lambda}_{k0}$ and $\widetilde{\lambda}_{k1}$ are exchanged. It is clear that applying the agent's map twice will have the same effect as applying the DEJMPS protocol twice, which means that for an even number of recurrence step they are equivalent.

As a side note, the other protocol that was found by the agent as described in the main text, applies such an additional operation before each entanglement purification step as well: Applying $H \otimes H$ on $\rho$ exchanges $\lambda_{10}$ and $\lambda_{01}$. This also yields a successful entanglement purification protocol, however with a slightly worse performance.

\subsection{Automatic depolarization variant}

We also investigated a variant where after each purification step, the state is automatically depolarized before the protocol is applied again. That means if the first step brought the state up to the new fidelity $F^\prime$ it is then brought to the form: $F^\prime \Ketbra{\Phi^+}{\Phi^+} + \frac{1-F^\prime}{3} \left(\Ketbra{\Psi^+}{\Psi^+} + \Ketbra{\Phi^-}{\Phi^-} + \Ketbra{\Psi^-}{\Psi^-}\right)$. This can always be achieved without changing the fidelity \cite{bbpssw}.

\begin{figure}
    \centering
    \includegraphics[width=0.7\linewidth]{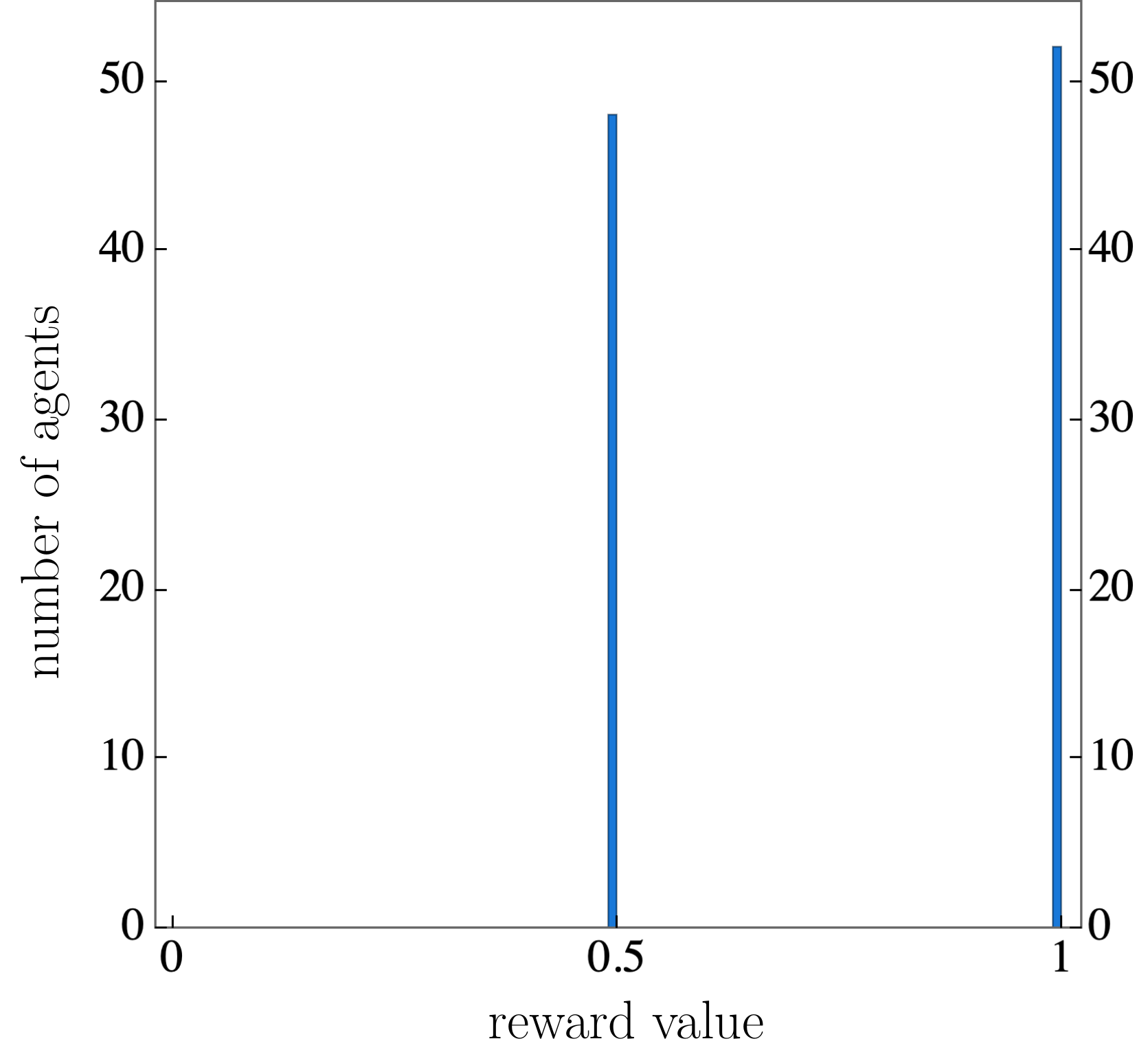}
    \caption{Entanglement purification environment with automatic depolarization after each purification step. Figure shows obtained rewards by 100 agents for their protocols found after $5 \times 10^5$ trials.}
    \label{fig:epp_depolarized_hist}
\end{figure}

In Fig.~\ref{fig:epp_depolarized_hist} the obtained reward for 100 agents for this alternative scenario is shown. The successful protocols consist of applying $\mathrm{CNOT}^{A_1 \rightarrow A_2} \otimes \mathrm{CNOT}^{B_1 \rightarrow B_2}$ followed by measuring qubits $A_2$ and $B_2$ in the computational basis. Optionally, some additional local operations that do not change the fidelity itself can be added as the effect of those is undone by the automatic depolarization. Similar to the scenario in the main text, there are some solutions that only accept one branch as successful, which means they only get half the reward as the success probability at each step is halved (center peak in Fig.~\ref{fig:epp_depolarized_hist}). The protocols for which two relevant branches are accepted are equivalent to the entanglement purification protocol presented in \cite{bbpssw}.

\section{Quantum repeater}

\subsection{Environment description}

The setup is depicted in Fig.~\ref{fig:repeater2_results}a. The repeater stations share entangled states via noisy channels with their neighbors, which results in pairs with an initial fidelity of $F=0.75$. The previous two protocols now are available as the elementary actions for this more complex scenario.

\textbf{Goal:} Entangled pair between the left-most and the right-most station with fidelity above threshold fidelity $F_\mathrm{th} = 0.9$.

\textbf{Actions:}
    \begin{itemize}
        \item Purify a pair with one entanglement purification step. (left pair, right pair, the long-distance pair that arises from entanglement swapping at the middle station)
        \item Entanglement swapping at the middle station
    \end{itemize}

We use the protocol in \cite{bbpssw} for this as it is computationally easier to handle. For a practical application it would be advisable to use a more efficient entanglement purification protocol.

\textbf{Percepts:} Current position of the pairs and fidelity of each pair.

\textbf{Reward function:} $R = \left(\frac{\textrm{const}}{\textrm{resources}} \right)^2$. The whole path is rewarded in full. The reward constant is obtained from an initial guess using the working-fidelity strategy described in \ref{sec:wfstrategy}.

\section{Scaling repeater}

\begin{figure*}[ht!]
    \centering
    \includegraphics[width=1\linewidth]{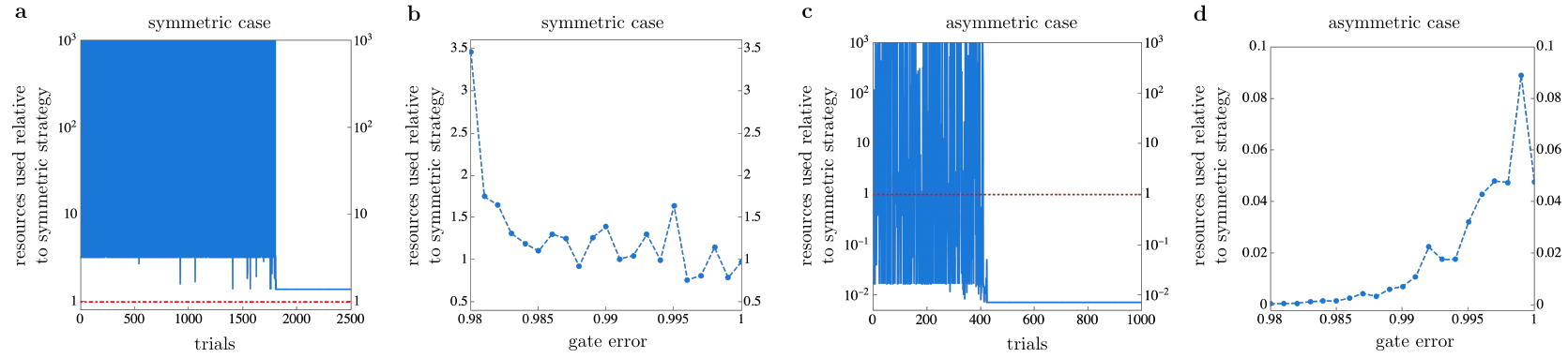}
    \caption{\textbf{a-b} Scaling repeater with 8 repeater links with symmetric initial fidelities of $0.7$. \textbf{a} Best solution found by an agent for gate reliability $p=0.99$. \textbf{b} Relative resources used by the agent's solution compared to the working-fidelity strategy for different gate reliability parameters. \textbf{c-d} Scaling repeater with 8 repeater links with very asymmetric initial fidelities (0.95, 0.9, 0.6, 0.9, 0.95, 0.95, 0.9, 0.6). \textbf{c} Best solution found by an agent for gate reliability $p=0.99$. \textbf{d} Relative resources used by the agent's solution compared to the working-fidelity strategy for different gate reliability parameters.}
    \label{fig:scaling_delegated_more}
\end{figure*}

\subsection{Environment description}

In addition to the elementary actions from the distance-2 quantum repeater discussed above we provide the agent with the ability to delegate solving smaller-scale problems of the same type to other agents, therefore splitting the problem into smaller parts. Then, the found sequence of actions is applied as one \textit{block action} as illustrated in Fig.~\ref{fig:scaling_repeater}a.

\textbf{Goal:} Entangled pair between the left-most and the right-most station with fidelity above threshold fidelity $F_\mathrm{th} = 0.9$.

\textbf{Actions:}
    \begin{itemize}
        \item Purify a pair with one entanglement purification step.
        \item Entanglement swapping at a station
        \item \textit{Block actions} of shorter lengths
    \end{itemize}

So for the setup with $L$ repeater links, initially there are $L$ purification actions and $L-1$ entanglement swapping actions. Of course, the purification actions have to be adjusted every time an entanglement swapping is performed to include the new, longer-distance pair. The block actions can be applied at different locations, e.g. example a length two block action can initially be applied at $L-1$ different positions (which also have to be adjusted to include longer-distance pairs as entanglement swapping actions are chosen). So it is easy to see how the action space quickly gets much larger as $L$ increases.

\textbf{Percepts:} Current position of the pairs and fidelity of each pair.

\textbf{Reward:} Again we use the resource-based reward function as this is the metric we would like to optimize. $R = \left(\frac{\textrm{const}}{\textrm{resources}} \right)^2$. The whole path is rewarded in full. The reward constant is obtained from an initial guess (see \ref{sec:wfstrategy}) and adjusted downward once a better solution is found such that the maximum possible reward from one trial is $1$.

\textit{Comment on block actions:} The main agent can use block actions for a wide variety of situations at different stages of the protocol. This means the sub-agents are tasked with finding block actions for a wide variety of initial fidelities, so a new problem needs to be solved for each new situation. In order to speed up the trials we save situations that have already been solved by sub-agents in a big table and reuse the found action sequence if a similar situation arises.

\subsubsection{Symmetric variant}
We force a symmetric protocol by modifying the actions as follows:

\textbf{Actions:}
    \begin{itemize}
        \item Purify all pairs with one entanglement purification step.
        \item Entanglement swapping at every second active station
        \item \textit{Block actions} of shorter lengths, that have been obtained in the same, symmetrized manner.
    \end{itemize}

\subsection{Additional results and discussion}

We also investigated different starting situations for this setup. Here we discuss two of them:

First, we also applied the agent that is not restricted to symmetric protocols to a symmetric starting situation. The results for initial fidelities $F = 0.7$ can be found in Fig.~\ref{fig:scaling_delegated_more}a-b. In general the agent finds solutions that are very close but not equal to the working-fidelity strategy described in \ref{sec:wfstrategy}. Remarkably, for some reliability parameters $p$ the agent even finds a solution that is slightly better by switching around the order of operations a little, or a threshold effect, where omitting an entanglement purification step on one of the pairs is still enough to reach the desired threshold fidelity.

Finally, we also looked at a situation that is highly asymmetric with starting fidelities (0.95, 0.9, 0.6, 0.9, 0.95, 0.95, 0.9, 0.6). Thus there are high-quality links on most connections, but two links suffer from very high levels of noise. The results depicted in Fig.~\ref{fig:scaling_delegated_more}c-d show that the advantage over a working-fidelity strategy is even more pronounced.

\subsection{Repeater stations for setups with memory errors \label{sec:appendix_repeater_positions}}
As mentioned in the main text, in order to properly account for imperfections in the quantum memories, we need to know the distance between repeater stations. 

In section \ref{sec:imperfect_memories} we looked at a total distance of just below $78.9 \text{\ km}$ with 7 intermediate repeater stations located at the positions shown in Table \ref{tab:repeater_positions}. Together with the distance dependent error model introduced in that section, these give rise to the asymmetric initial fidelities (0.8,0.6,0.8,0.8,0.7,0.8,0.8,0.6) which we also used for the setup with perfect memories.

In section \ref{sec:finding_positions} we considered a list of 7 possible locations to position repeater stations on, which we get by scaling down the previous scenario to a total distance of $20 \text{\ km}$. We chose such a comparably short distance in order to make protocols with only one added repeater station a viable solution. The positions of the 7 possible locations are listed in Table \ref{tab:possible_locations}.

\begin{table}[]
    \centering
    \begin{tabular}{c|c}
         repeater station index & position \\
         \hline
         1 & $6.8 \text{\ km}$\\
         2 & $23.6 \text{\ km}$\\
         3 & $30.4 \text{\ km}$\\
         4 & $37.2 \text{\ km}$\\
         5 & $48.5 \text{\ km}$\\
         6 & $55.3 \text{\ km}$\\
         7 & $62.1 \text{\ km}$
    \end{tabular}
    \caption{The positions of the repeater stations in section \ref{sec:imperfect_memories}. The terminal stations between which shared entanglement is to be established are located at positions $0 \text{\ km}$ and $78.9 \text{\ km}$.}
    \label{tab:repeater_positions}
\end{table}

\begin{table}[]
    \centering
    \begin{tabular}{c|c}
         possible location index & position \\
         \hline
         1 & $1.73 \text{\ km}$\\
         2 & $5.98 \text{\ km}$\\
         3 & $7.71 \text{\ km}$\\
         4 & $9.44 \text{\ km}$\\
         5 & $12.29 \text{\ km}$\\
         6 & $14.02 \text{\ km}$\\
         7 & $15.75 \text{\ km}$
    \end{tabular}
    \caption{Possible locations at which repeater stations can be positioned in section \ref{sec:finding_positions}. The terminal stations between which shared entanglement is to be established are located at positions $0 \text{\ km}$ and $20 \text{\ km}$.}
    \label{tab:possible_locations}
\end{table}

\subsection{Working-fidelity strategy \label{sec:wfstrategy}}

This is the strategy we use to determine the reward constants for the quantum repeater environments and was presented in \cite{Br98}. This strategy leads to a resource requirement per repeater station that grows logarithmically with the distance.

For repeater lengths with $2^k$ links it is a fully nested scheme and can therefore be stated easily:
\begin{enumerate}
    \item Pick a working fidelity $F_w$.
    \item Purify all pairs until their fidelity is $F \geq F_w$.
    \item Perform entanglement swapping at every second active station such that there are half as many repeater links left.
    \item Repeat from step 2. until only one pair remains (and therefore the outermost stations are connected).
\end{enumerate}
We then optimize the choice of $F_w$ such that the resources are minimized for the given scenario.

As we are dealing with repeater lengths that are not a power of $2$ as part of the delegated subsystems discussed in the main text, the strategy is adjusted as follows for those cases.
\begin{enumerate}
    \item Pick a working fidelity $F_w$.
    \item Purify all pairs until their fidelity is $F \geq F_w$.
    \item Perform entanglement swapping at the station with the smallest combined distance of their left and right pair (e.g. 2 links + 3 links). If multiple stations are equal in that regard, pick the leftmost station.
    \item Repeat from step 2. until only one pair remains (and therefore the outermost stations are connected).
\end{enumerate}
Then, we again optimize the choice of $F_w$ such that the resources are minimized for the given scenario.

\section{Computational Resources\label{sec:runtimes}}
All numerical calculations were performed on a machine with four Sixteen-Core AMD Opteron 6274 CPUs. In Table \ref{tab:runtimes} we provide the run-times of the calculations presented in this paper. Memory requirements were insignificant for all of our scenarios.

\begin{table*}[]
    \centering
    \begin{tabular}{c|c|c}
         scenario & run-time\\
         \hline
         Teleportation (Fig. \ref{fig:teleportation_results}b/c )& Clifford: $\sim$5h, Universal: $\sim$3d \\
         Teleportation Variant 1 & $<$1h \\
         Teleportation Variant 2 (Fig. \ref{fig:teleportation_results}e) & $\sim$93h \\
         Entanglement Purification (Fig. \ref{fig:epp_results}) & $\sim$7h\\
         Quantum Repeater (Fig. \ref{fig:repeater2_results}) & $<$1h \\
         Scaling symmetric case (Fig. \ref{fig:scaling_repeater} b, c) & $\sim$3h \\
         Scaling asymmetric case (Fig. \ref{fig:scaling_repeater} d, e) & $\sim$24h per data point \\
         Scaling with memory errors (Fig. \ref{fig:fig6_memory_times}) & $\sim$25h per data point \\
         Repeater station positions (Table \ref{tab:positions_table}) & $\sim$4.5h 
    \end{tabular}
    \caption{Run time for our scenarios on a machine with four Sixteen-Core AMD Opteron 6274 CPUs.}
    \label{tab:runtimes}
\end{table*}

\end{document}